\documentclass{PoS}

\title{The H.E.S.S. Gravitational Wave Rapid Follow-up Program during O2 and O3}

\ShortTitle{The H.E.S.S. GW follow-up program}

\author{\speaker{Halim Ashkar}\\
        {$^1$}IRFU, CEA, Universit\'e Paris-Saclay, F-91191 Gif-sur-Yvette, France\\
        E-mail: \email{halim.ashkar@cea.fr}}
        
\author{Francois Brun$^1$, Matthias F\"u{\ss}ling$^2$, Clemens Hoischen$^3$, Stefan Ohm$^2$, Heike Prokoph$^2$, Patrick Reichherzer$^{1,4,5}$, Fabian Sch\"ussler$^1$, Monica Seglar Arroyo$^{6}$\\
{$^2$}DESY, D-15738 Zeuthen, Germany\\
{$^3$}Institut f\"ur Physik und Astronomie, Universit\"at Potsdam, Potsdam, Germany\\
{$^4$}Ruhr-Universität Bochum, Universitätsstraße 150, 44801 Bochum, Germany\\
{$^5$}Ruhr Astroparticle and Plasma Physics Center, Ruhr-Universität Bochum, 44780 Bochum, Germany\\
{$^6$}Univ. Savoie Mont Blanc, CNRS,
Laboratoire d’Annecy de Physique des Particules - IN2P3,
74000 Annecy, France}

\abstract{Since 2015, the direct detection of Gravitational Waves (GWs) became possible with ground-based interferometers like LIGO and Virgo. GWs became the center of attention of the astronomical community and electromagnetic observatories took a particular interest in follow-up observations of such events. The main setback of these observations is the poor localization of GW events. In fact, GW localization uncertainties can span tens to hundreds of deg$^{2}$ the sky even with the advanced configurations of current GW interferometers. In this contribution, we present five follow-up strategies developed for the High Energy Stereoscopic System (H.E.S.S.) and assess their performances. We show how a 2D and 3D galaxy targeted search approach exploiting the integral probability inside the instruments field of view are best suited for medium field of view instruments like H.E.S.S. We also develop an automatic response scheme within the H.E.S.S. Transient Follow-up system that is optimized for fast response and is capable of responding promptly to all kind of GW alerts. GW events are filtered by the developed scheme and prompt and afterglow observations are automatically scheduled. The H.E.S.S. response latency to prompt alerts is measured to be less than 1 minute. With this continually optimized GW response scheme, H.E.S.S. scheduled several GW follow-up observations during the second and third LIGO/Virgo observation runs.}

\FullConference{37th International Cosmic Ray Conference -ICRC2021-\\
		July 12th - July 23rd, 2021\\
		Berlin, Germany}

\begin{document}

\section{Introduction}
The detection of the Binary Neutron Star (BNS) merger GW170817~\cite{GW170817} and its Gamma-ray Burst (GRB) counterpart~\cite{GW_GRB_170817} triggered the most extensive multi-wavelength follow-up campaign known to our days~\cite{GW170817_MM}. Gravitational Wave (GW) events continue to spark much interest in the astrophysical community and the search for electromagnetic (EM) counterparts continue to be part of the challenges during GW observing runs. For the Very High Energy (VHE) gamma-ray domain, several models predict long lasting emission in a short GRB form (e.g.~\cite{Murase_2017}~\cite{Veres})

However, GW events suffer from poor localisation and their uncertainty regions can span tens to hundreds of degrees in the sky. Therefore in order to efficiently follow a GW event, special "tiling" strategies are used in order to maximize the chance of observing an EM counterpart.

The High Energy Stereoscopic System (H.E.S.S.) is composed of five Imaging  Atmospheric Cherenkov Telescopes (IACTs) (four 12-m and one 28-m telescopes) and is sensitive to VHE gamma rays with energies of few tens of GeV up to 100 TeV. It is located in the Khomas Highland in Namibia more than 100 km away from the capital and minimally affected by its light pollution. The 12-m telescopes have a $\sim$2.5 deg radius Field of view (FoV) and the 28-m telescope has a $\sim$1.5 deg radius FoV. H.E.S.S. dedicates a part of its observing program to the follow-up of GW events. For this purpose special strategies have been developed in the collaboration and a dedicated GW follow-up program has emerged. 

In this contribution, the H.E.S.S. rapid GW follow-up program is presented. This program handles rapid responses to GW events in the scale of few hours to days. For long-term follow-ups, involving the re-brightening of the afterglow phase due to the viewing angle, observation campaigns are triggered based on multi-wavelength (MWL) information monitoring. Such triggers are not described in this contribution and an example can be found in~\cite{EM170817_HESS}.

The paper is organized as follows: in Sec.~\ref{sec:overview_GW_IACTs} an overview of GW follow-ups with IACTs is given, in Sec.~\ref{sec:followup_strategies} five follow-up strategies used within the H.E.S.S. collaboration are described. Finally, in Sec.~\ref{sec:GW_MODULE} the H.E.S.S. GW module that allows for automatic response to GW alerts is presented and we conclude at the end of this section.

\section{GW follow-up strategies with IACTs}
\label{sec:overview_GW_IACTs}
Three main ingredients are taken into consideration to build the H.E.S.S. GW follow-up strategies: the GW localisation maps provided by the Ligo and Virgo Collaboration (LVC), the telescope observation constraints and external information like galaxy catalogs.  
\begin{itemize}
    \item GW localisation maps are provided in HEALPix format~\cite{healpix}. They usually contain four layers of information, one layer containing the posterior probability that the source is contained inside a pixel $i$, $\rho_i$ and the remaining three layers contain event distance information~\cite{singer2016going}. The resolution of the map is defined by the \textit{N$_{side}$} parameter and the total number of pixels of the map is $N_{pix}=12\times N^2_{side}$
    \item Telescope observation condition can be devided into two categories, observation constraints and visibility constraints. The visibility of the telescope is defined by the maximum allowed zenith angle. In fact, for IACTs the energy threshold of observations range from a few tens to hundreds of GeVs, depending on the zenith angle under which the source is observed. Therefore H.E.S.S. observations at low zenith angles, are preferred and the maximum allowed zenith angle is $\theta{z} < 60 \, \mathrm{deg}$. Observation constraints are defined by the conditions of observation that require low level of background light falling in the cameras photomultipliers. H.E.S.S. observations were typically restricted to astronomical darkness with the Moon below the horizon. However, H.E.S.S. is also capable of observing with low levels of moonlight. This is achieved by monitoring the moon's phase, altitude and moon to source separation. We also note that H.E.S.S. observations have a standard 28 minute duration. But this value can be changed given the available observation time during a night.
    \item External information like the distribution of the galaxies in the local Universe can be used in case distance information from the GW maps are used. These information are provided by galaxy catalogs. The Advanced Detector Era (GLADE) catalog~\cite{dalya2018glade} is currently being used in the H.E.S.S. GW program due to its easy availability and continuous updating. Moreover, additional information like galaxy stellar masses are also considered by using the MANGROVE~\cite{MANGROVE} catalog. 
\end{itemize}
GW follow-up algorithms will assess the available observation time during a given astronomical night, divide this time into observation windows and find the best position to observe during each observation window. The night is divided into observation windows, making sure that the minimum window duration is 10 minutes. The steps of the procedure are as follows:
\begin{enumerate}
\item Select the most probable sky location fulfilling the IACT observation conditions (e.g. zenith angle range, dark time, etc.)
\item Schedule observation for this direction at a time $T_i$, where $i=0$ for the first observation, with a duration $\Delta$t.
\item Mask a circular sky region representing the effective IACT FoV around that region.
\item Using the modified visibility window $T_i =T_0 +  i \cdot \Delta$t, where $i$ is the observation number, and the iteratively masked skymap in its new position in the sky at $T_i$, steps 1-3 are repeated until $\gamma$-ray emission is detected by the real-time analysis, the covered probability for the next observations is insignificant or the allocated observation time is used (end of the Target of Opportunity (ToO) observations). 
\end{enumerate}

To find the best position to observe at a time $T_i$, several strategies are considered. These strategies are presented in the following section.

\section{H.E.S.S. GWs follow-up strategies}
\label{sec:followup_strategies}
GW follow-up strategies either rely solely on the posterior probability contained inside the first layer of each pixel (2D follow-up) of the GW localisation map, or take into consideration the distribution of the galaxies in the local Universe and convolute it with the probability and distance information from the GW map contained inside the four layers of the pixels (3D follow-up) in order to reduce the search to few galaxies instead of entire region in the sky.

\begin{figure}[hb!]
  \centering
  \begin{minipage}[b]{0.24\textwidth}
    \includegraphics[width=\textwidth]{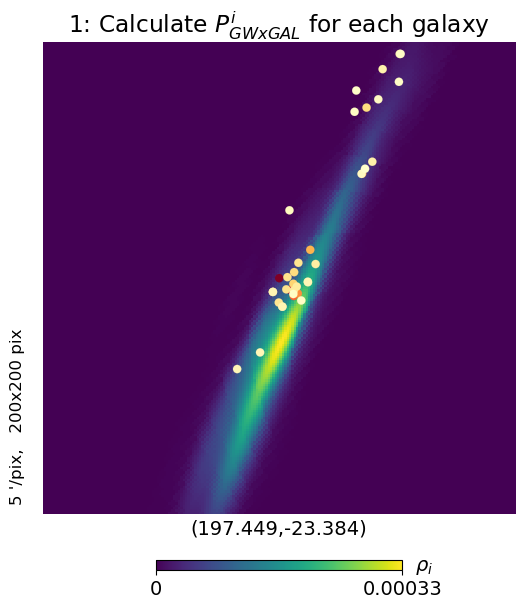}
  \end{minipage}
  \hfill
  \begin{minipage}[b]{0.24\textwidth}
    \includegraphics[width=\textwidth]{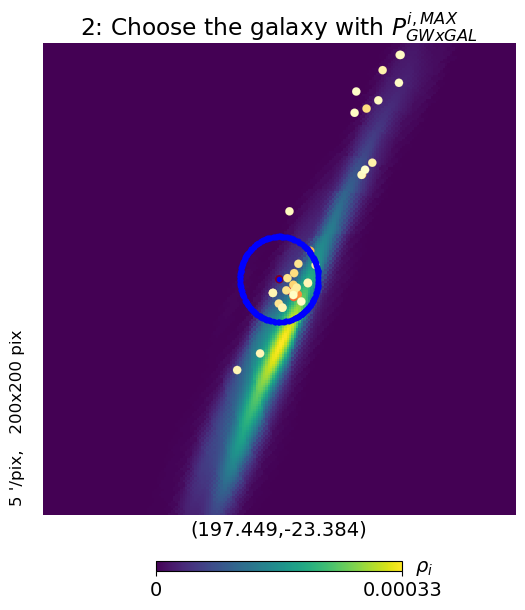}
    \end{minipage}
      \hfill
      \begin{minipage}[b]{0.24\textwidth}
    \includegraphics[width=\textwidth]{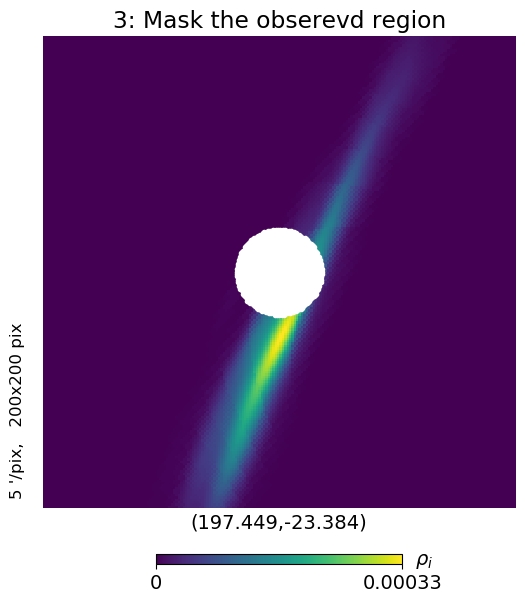}
    \end{minipage}
      \hfill
          \begin{minipage}[b]{0.24\textwidth}
    \includegraphics[width=\textwidth]{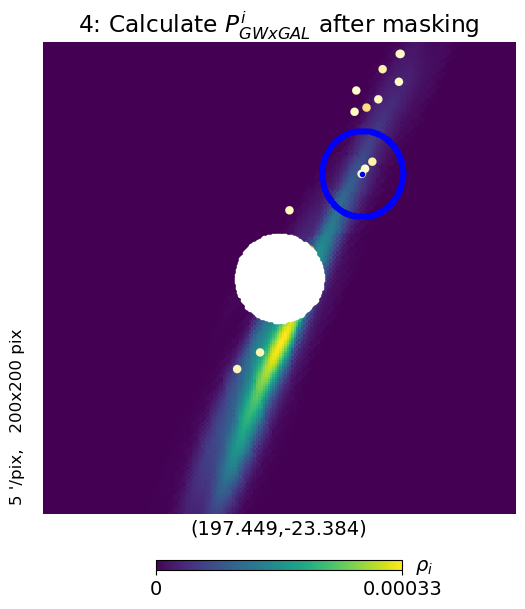}
    \end{minipage}
  \caption{Graphical representation of the steps to compute the best observable position for two windows for a galaxy-targeted 3D search. These steps are repeated to determine the best position to observe for each observation window. The \textit{updated} map for GW170817 is used. The colored dots represent the galaxies with the highest $P^i_{GW \times GAL}$  in the region on the YlOrRd color scale~\cite{matplotlib}. The blue dot represent the galaxy that has the highest probability to be observed and chosen for observations. The blue circle represent the H.E.S.S. FoV. The white region is masked.}
  \label{fig:BestGal_steps}
\end{figure}

\begin{figure}[hb!]
  \centering
  \begin{minipage}[b]{0.24\textwidth}
    \includegraphics[width=\textwidth]{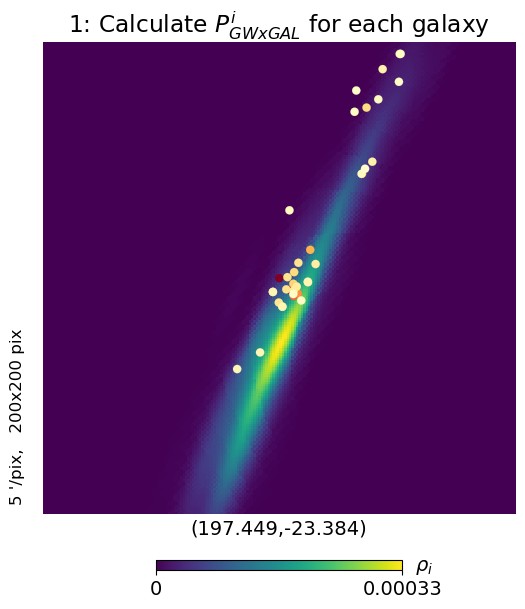}
  \end{minipage}
  \hfill
  \begin{minipage}[b]{0.24\textwidth}
    \includegraphics[width=\textwidth]{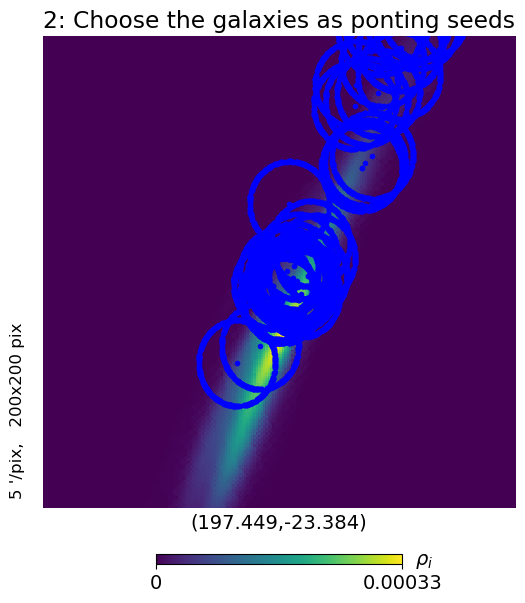}
    \end{minipage}
      \hfill
      \begin{minipage}[b]{0.24\textwidth}
    \includegraphics[width=\textwidth]{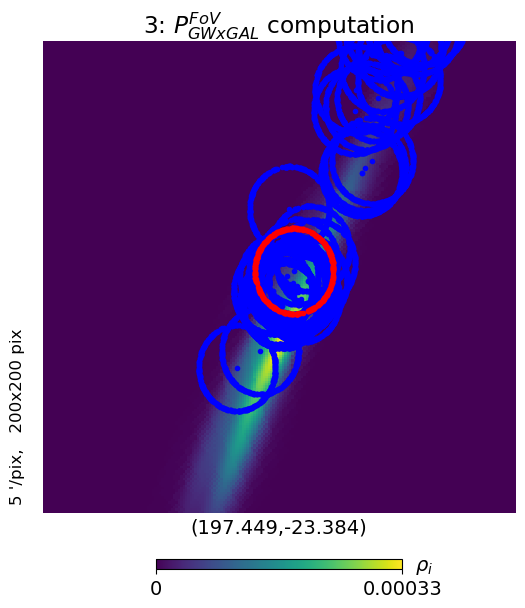}
    \end{minipage}
      \hfill
          \begin{minipage}[b]{0.24\textwidth}
    \includegraphics[width=\textwidth]{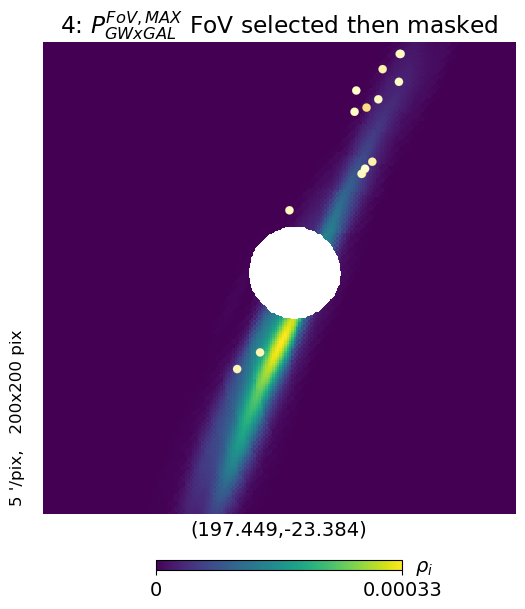}
    \end{minipage}
  \caption{Graphical representation of the steps to compute the best observable position for one window for the FoV-targeted 3D search with galaxies as seeds. These steps are repeated to determine the best position to observe for each observation window. The \textit{updated} map for GW170817 is used. The blue dots represent the centers of the IACT FoVs inside which $P^{FoV,i}_{GW \times GAL}$ will be calculated.  The blue circles are the considered FoVs. The red circle represents the region with $P^{FoV,MAX}_{GW \times GAL}$. The white region is masked. The colored dots represent the galaxies with the highest $P^i_{GW \times GAL}$  in the region on the YlOrRd color scale~\cite{matplotlib}.}
  \label{fig:PalinFoV_steps}
\end{figure}

\begin{figure}[hb!]
  \centering
  \begin{minipage}[b]{0.24\textwidth}
    \includegraphics[width=\textwidth]{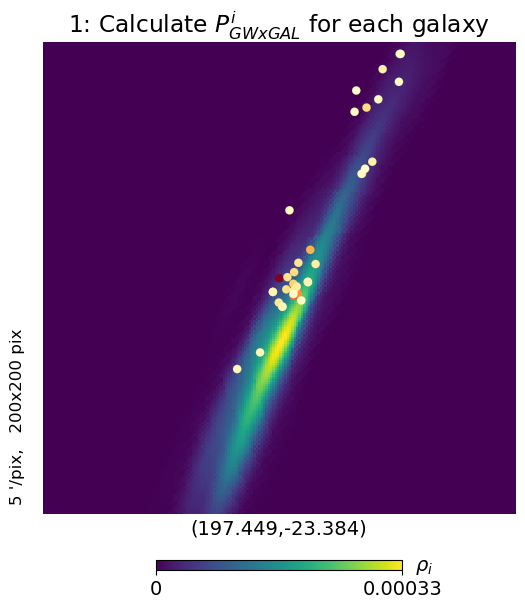}
  \end{minipage}
  \hfill
  \begin{minipage}[b]{0.24\textwidth}
    \includegraphics[width=\textwidth]{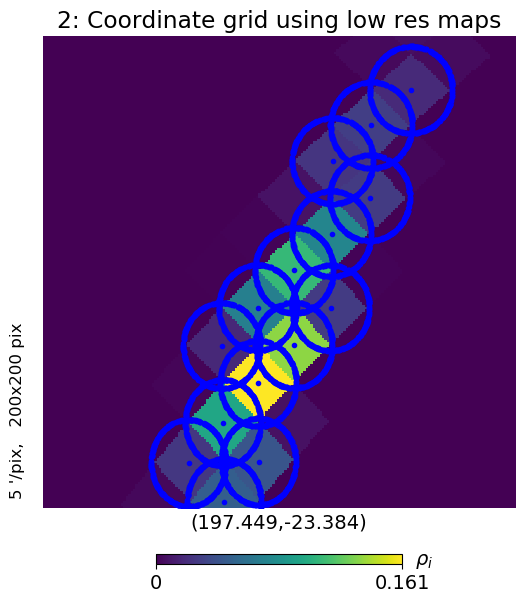}
    \end{minipage}
      \hfill
     \begin{minipage}[b]{0.24\textwidth}
    \includegraphics[width=\textwidth]{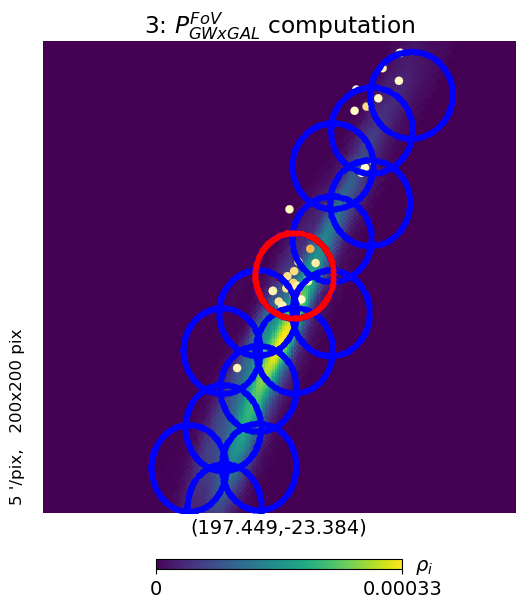}
  \end{minipage}
  \hfill
  \begin{minipage}[b]{0.24\textwidth}
    \includegraphics[width=\textwidth]{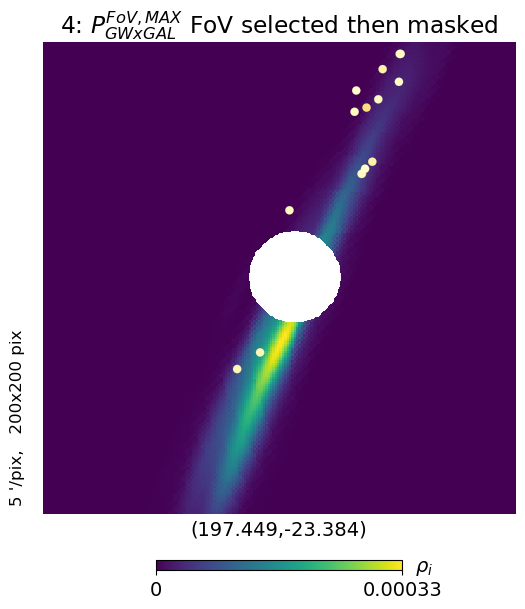}
  \end{minipage}
  \caption{Graphical representation of the steps to compute the best observable position for one window for the FoV-targeted 3D search with a coordinate grid. These steps are repeated to determine the best position to observe for each observation window. The \textit{updated} map for GW170817 is used. For representation purposes the region enclosed in the  90\% localisation uncertainty and $N_{side} = 32$ is chosen for the construction of the low resolution coordinate grid. The blue dots represent the grid of coordinates of the IACT FoVs inside which $P^{FoV}_{GW \times GAL}$ will be calculated.  The blue circles are the considered FoVs. The red circle represents the region with $P^{FoV,MAX}_{GW \times GAL}$. The white region is masked. The colored dots represent the galaxies with the highest $P^i_{GW \times GAL}$  in the region on the YlOrRd color scale~\cite{matplotlib}. From~\cite{Technical_paper}.}
  \label{fig:PGal_PR_steps}
\end{figure}

We define the following entities: 
\begin{itemize}
    \item $\rho_i$ is the probability of containing the GW event for each pixel $i$.
    \item $P^{FoV}_{GW}$ is the integrated probability inside the H.E.S.S. FoV
    \begin{equation}
    P^{FoV,i}_{GW}=\int^{2\pi}_{0}\int_0^{r_{FoV}}\rho(r,\phi) \,\mathrm{d}r \mathrm{d}\phi
    \label{eq:PGWFOV}
    \end{equation}
    where $r$ is the radius and $\phi$ the angle of the FoV represented by a circle.
    \item $P^{FoV,Max}_{GW}$ is the maximum achievable integrated probability inside the H.E.S.S. FoV for a given time.
    \item $P^i_{{GWxGAL}}$ is the probability assigned to the galaxies by correlating the GW map and the galaxy catalogs. With
    
\begin{equation}
\frac{\mathrm{d}P}{\mathrm{d}V} = \rho_i \frac{N_\mathrm{pix}}{4\pi} \frac{\hat{N_i}}{\sqrt{2\pi}\hat{\sigma}_i}\text{exp}\bigg[ -\frac{(z-\hat{\mu_i})^2}{2\hat{\sigma}_i^2} \bigg] 
\label{eq3}
\end{equation}

where $\hat{\mu}_i$, $\hat{\sigma}_i$ and $\hat{N}_i$ refer to the mean, the scale, and the normalization and

\begin{equation}
P^i_{GW \times GAL}=\frac{\mathrm{d}P^i/\mathrm{d}V}{\sum_j \mathrm{d}P^j/\mathrm{d}V}
\label{eq4}
\end{equation}
Additional galaxy parameters like the stellar mass~\cite{MANGROVE} can be taken into consideration in this step.
    \item $P^{FoV}_{GW \times GAL}$ is the integrated galaxy probability inside the H.E.S.S. FoV
    
    \begin{equation}
P^{FoV,i}_{GW \times GAL}=\int^{2\pi}_{0}\int_0^{r_{FoV}}P^i_{GW \times GAL}(r,\phi)\, \mathrm{d}r \mathrm{d}\phi
\label{EqGal}
\end{equation}

    \item $P^{FoV,MAX}_{GW \times GAL}$ is the maximum achievable integrated galaxy probability inside the H.E.S.S. FoV for a given time.
\end{itemize} 
The following figures (Fig.~\ref{fig:PBest_steps},~\ref{fig:PGW_steps},~\ref{fig:BestGal_steps},~\ref{fig:PalinFoV_steps} and~\ref{fig:PGal_PR_steps}) describe the five GW follow-ups strategies developed within the H.E.S.S. collaboration. The strategies are integrated in the H.E.S.S. GW framework; Depending on the GW alert parameters and properties, a strategy can be chosen to schedule H.E.S.S. follow-up observations as described in the next section.  
 
\newpage
\begin{figure}[htb!]
  \centering
  \begin{minipage}[b]{0.24\textwidth}
    \includegraphics[width=\textwidth]{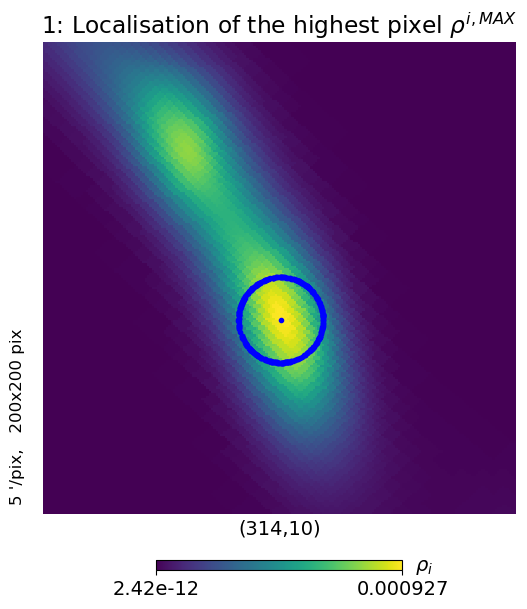}
  \end{minipage}
  \hfill
  \begin{minipage}[b]{0.24\textwidth}
    \includegraphics[width=\textwidth]{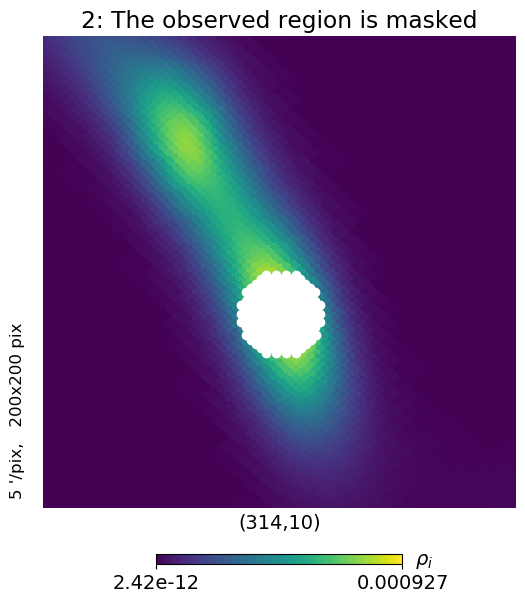}
    \end{minipage}
      \hfill
     \begin{minipage}[b]{0.24\textwidth}
    \includegraphics[width=\textwidth]{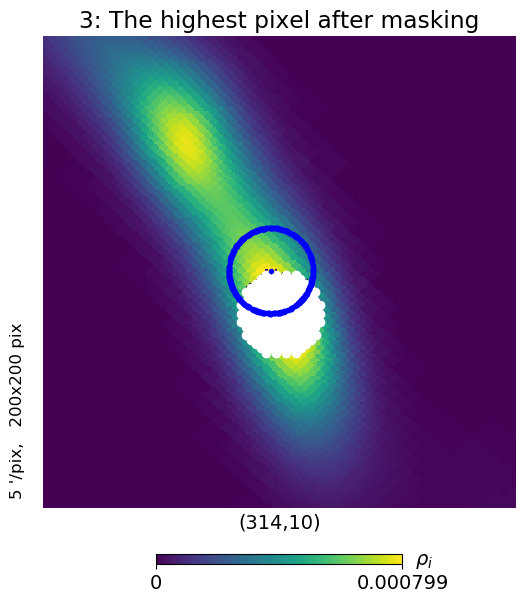}
  \end{minipage}
  \caption{Graphical representation of the steps to compute the best observable position for two consecutive time window for a pixel-targeted 2D search. These steps are repeated to determine the best position to observe for each observation window. The \textit{initial} map for the GW event S190728q is used as an example. For representation purposes $N_{side}$ has been reduced to  $N_{side} =2 56$ in order to be able to see the pixels. The blue dot represents the location of the pixel with highest probability $\rho_{i}$. The blue circles are the considered FoVs. The white region is masked.}
  \label{fig:PBest_steps}
\end{figure}
\begin{figure}[hb!]
  \centering
  \begin{minipage}[b]{0.24\textwidth}
    \includegraphics[width=\textwidth]{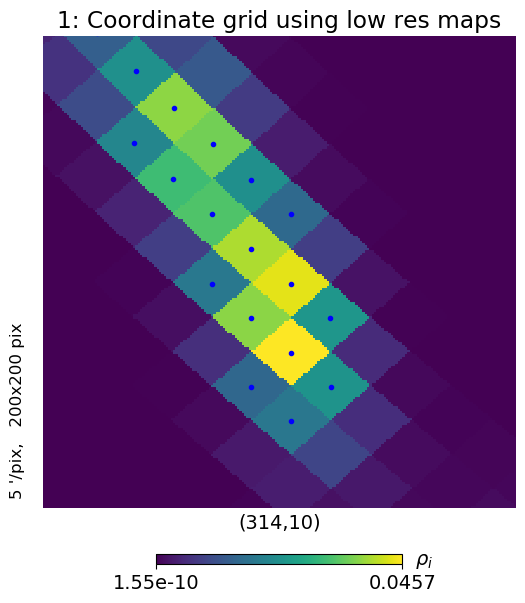}
  \end{minipage}
  \hfill
  \begin{minipage}[b]{0.24\textwidth}
    \includegraphics[width=\textwidth]{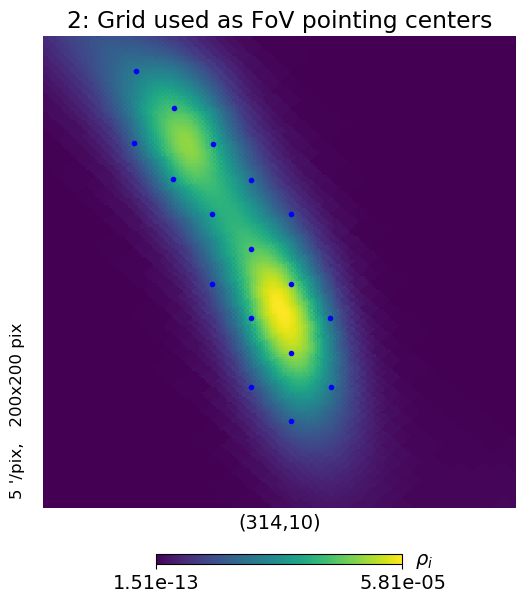}
    \end{minipage}
      \hfill
     \begin{minipage}[b]{0.24\textwidth}
    \includegraphics[width=\textwidth]{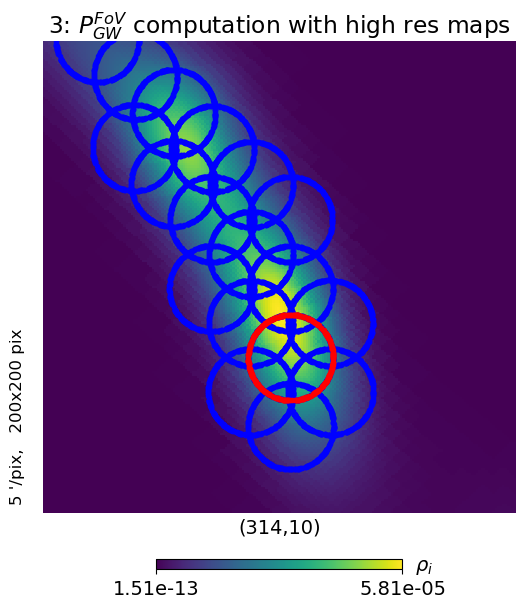}
  \end{minipage}
  \hfill
  \begin{minipage}[b]{0.24\textwidth}
    \includegraphics[width=\textwidth]{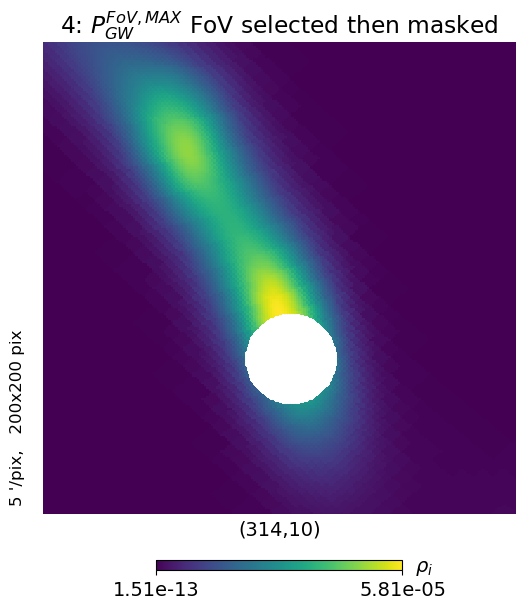}
  \end{minipage}
  \caption{Graphical representation of the steps to compute the best observable position for one window for a FoV-targeted 2D search. These steps are repeated to determine the best position to observe for each observation window. The \textit{initial} map for the GW event S190728q is used as an example. For representation purposes the region enclosed in the  50\% localisation uncertainty and $N_{side} =32$ are chosen for the construction of the low resolution coordinate grid. The blue dots represent the grid of coordinates of the IACT FoVs inside which $P^{FoV}_{GW}$ will be calculated. The blue circles are the considered FoVs. The red circle represents the region with $P^{\text{FoV},MAX}_{GW}$. The white region is masked. From~\cite{Technical_paper}.}
  \label{fig:PGW_steps}
\end{figure}

\newpage

\newpage

\section{The H.E.S.S. automatic GW module}
\label{sec:GW_MODULE}

The H.E.S.S. ToO Alert System is part of the H.E.S.S. Transients Follow-up System described in~\cite{VoAlerter_pub} and~\cite{Technical_paper}. It is responsible for handling ToO alerts by assessing their potential interest to H.E.S.S., their visibility and the possibility of follow-up observations. GW alerts are received through the The Gamma-ray Coordinates Network (GCN) network~\cite{GCN}. Four types of GW alerts are currently available for distribution: \textit{early-warning}, \textit{preliminary}, \textit{initial} and \textit{update}. Details on these alerts can be found in~\cite{Public_alert_guide}.

\begin{figure}
  \centering
\includegraphics[width=0.95\textwidth]{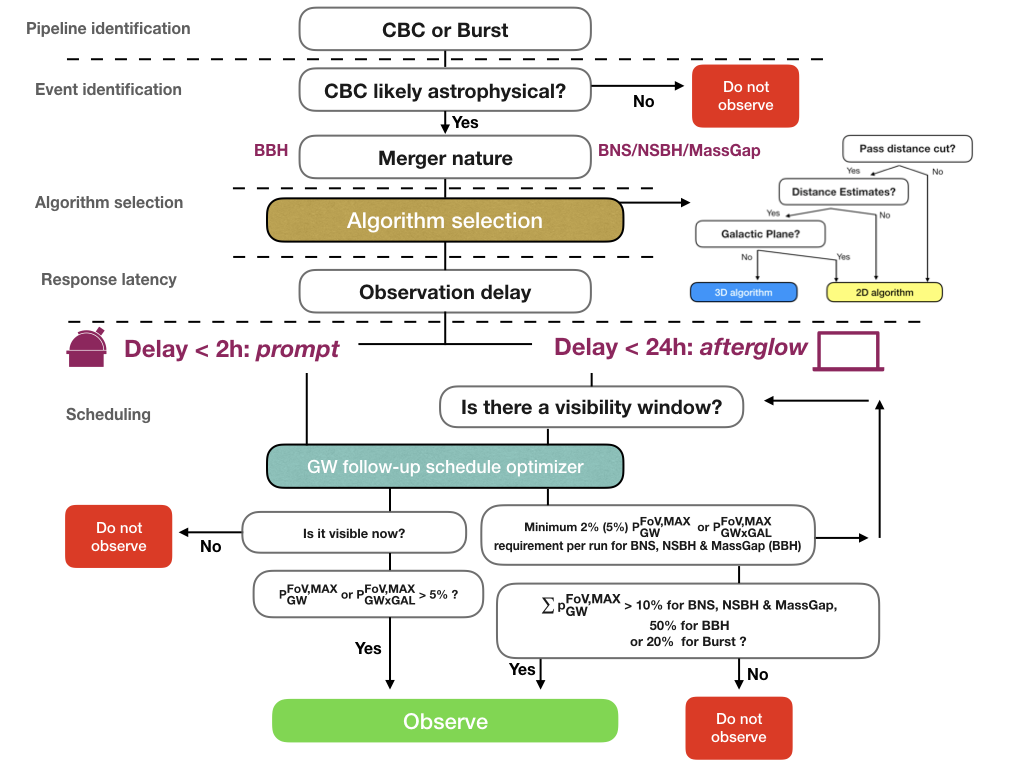}
\caption{Schematic overview of the decision tree used in the automatic response of H.E.S.S to GW events.}
\label{fig:GWVoAlerter}
\end{figure}

The H.E.S.S. GW module is integrated in the H.E.S.S. ToO Alert System and is illustrated in Fig.~\ref{fig:GWVoAlerter}. The GW module handles all four types of GW alerts. Upon the reception of an alert the GW module parses the information from the alert on the trigger and makes the decision to follow or not based on the criteria shown in Fig.~\ref{fig:GWVoAlerter}. The used information are the time of the GW event and alert, the detecting pipeline, the alert classification (BNS or BBH) and the link to the GW localisation map.

Based on the 3D information (distance) contained in the alert the H.E.S.S. GW module chooses a 2D or a 3D strategy. If distance information are available, and the GW lies in a region where galaxy catalogs are complete (i.e. at small distance and outside the galactic plane) a 3D coverage is used. Otherwise, a 2D strategy is chosen. 

The module is divided into two parts, an \textit{afterglow} part and a \textit{prompt} part. In case an alert arrives during H.E.S.S. observing hours, the \textit{prompt} module immediately calculates the best position to be observed at the time of arrival of the alert, assesses its visibility and triggers automatic telescope response based on its assessment. Meanwhile, the \textit{afterglow} module computes the entire follow-up schedule for the night. This allows to save time. The prompt module does not handle high latency alerts (\textit{update}). If the alert arrives outside H.E.S.S. observing hours, only the \textit{afterglow} module is triggered. 

The strategies implemented in the system are the FoV-targeted 2D search for 2D follow-ups and the FoV-targeted 3D search with galaxies as seeds for 3D follow-ups. The decision to choose these strategies is based on the fact that the integration of the probability inside the H.E.S.S. medium FoV results in better coverage of the entire GW map during a night. Moreover, the FoV-targeted 3D search with galaxies as seeds has a very quick computation time for \textit{prompt} triggers. More details on this decision can be found in~\cite{Technical_paper}.

Finally, the latency of H.E.S.S. response is estimated to be less than 1 minute for most cases (including telescope slewing). This is the time it takes for H.E.S.S. to start data taking on the first observation position on a promising GW events followed with a 3D strategy.\\

In total, 11 GW events were detected during O1 and O2~\cite{O2_paper_catalog} and 80 alerts were sent during O3~\cite{O3a_paper_catalog}~\cite{GraceDB}. Of these 80 alerts, 24 were retracted, 3 were classified as cause from terrestrial noise, 52 were related to compact binary coalescences and 1 un-modelled burst alert. Most of the alerts could not be followed due to large localisation uncertainty and the constraint of the HESS observing committee to be able to cover a significant part of the region within one night (as indicated in Fig.~\ref{fig:GWVoAlerter}). From all the alert received during the three observing runs, H.E.S.S. triggered observations on 8 alerts. The well localised S190814bv could not be observed due to it falling during the full moon period. 2 of the follow-ups (S191204r and S200114f) could not be performed due to bad weather conditions. H.E.S.S. gathered data on 6 GW event: GW170814, GW170817, S190512at, S190728q, S200115j~\cite{S200115j} and S200224ca~\cite{S200224ca}. These observations are presented in~\cite{Technical_paper}. GW170817 was observed during a short-term and a long-term follow-up campaigns. The findings from these campaigns are presented in~\cite{GW170817_HESS} and~\cite{EM170817_HESS}. The analysis of the follow-up of the remaining events will be presented in dedicated publications~\cite{GWesultsHESS}~\cite{BBHHESS}.

\end{document}